\documentclass[12pt]{article}
\pdfoutput=1
\usepackage{putex}
\usepackage{feyn}
\usepackage[vcentermath]{youngtab}
\usepackage{subfig}
\usepackage{lscape}

\usepackage{graphicx}
\usepackage{epstopdf}
\usepackage{enumerate}
\usepackage{cite}
\usepackage{tensor}
\usepackage{slashed}
\usepackage{amsmath}
\usepackage{amssymb}
\usepackage{mathrsfs}
\usepackage{lgrind}

\usepackage{bbm}

\usepackage{hyperref}

\numberwithin{equation}{section}

\newcommand {\be} {\begin {equation}}
\newcommand {\ee} {\end {equation}}

\newcommand {\bes} {\begin {equation*}}
\newcommand {\ees} {\end {equation*}}

\newcommand{\eps}{\epsilon}

\def\CH{{\cal H}}

\newcommand{\beq}{\begin{equation}}
\newcommand{\eeq}{\end{equation}}

\def\be{ \begin{equation} }
\def\ee{ \end{equation} }

\begin{document}

\preprint{PUPT-2514}

\institution{PU}{Department of Physics, Princeton University, Princeton, NJ 08544}
\institution{PCTS}{Princeton Center for Theoretical Science, Princeton University, Princeton, NJ 08544}

\title{ Uncolored Random Tensors, Melon Diagrams, and the SYK Models
}

\authors{Igor R.~Klebanov\worksat{\PU,\PCTS} and Grigory Tarnopolsky\worksat{\PU}}

\abstract{Certain models with rank-$3$ tensor degrees of freedom have been shown by Gurau and collaborators
to possess a novel large $N$ limit, where
$g^2 N^3$ is held fixed. In this limit the perturbative expansion in the quartic coupling constant, $g$, is dominated by a special class of ``melon" diagrams.
We study ``uncolored" models of this type, which contain a single copy of real rank-$3$ tensor. Its three indices are distinguishable; therefore, 
the models possess $O(N)^3$ symmetry with the
tensor field transforming in the tri-fundamental representation. 
Such uncolored models
also possess the large $N$ limit dominated by the melon diagrams. The 
quantum mechanics of a real anti-commuting tensor therefore has a similar large $N$ limit to the model recently introduced by Witten as an implementation of
 the Sachdev-Ye-Kitaev (SYK) model which does not require disorder. 
Gauging the $O(N)^3$ symmetry in our quantum mechanical model removes the non-singlet states; therefore, one can search for its well-defined gravity dual. We point out, however,
that the model possesses a vast number of gauge-invariant operators involving higher powers of the tensor field, suggesting that the complete gravity dual will be intricate.
We also discuss the quantum mechanics of a complex 3-index anti-commuting tensor, which has $U(N)^2\times O(N)$ symmetry and argue that it is equivalent in the
large $N$ limit to a version of SYK model with complex fermions. 
Finally, we discuss similar models of a commuting tensor in dimension $d$. While the quartic interaction is not positive definite, we construct the large $N$ Schwinger-Dyson equation
for the two-point function and show that its solution is consistent with conformal invariance. We carry out a perturbative check of this result using the $4-\epsilon$ expansion.
}

\date{}
\maketitle

\tableofcontents

\section{Introduction}

An important tool in theoretical physics is the study of certain models in the limit where they have a large number of degrees of freedom.
Several different broad classes of such ``large $N$ limits" have been explored.  
Perhaps the most tractable large $N$ limit applies to
theories where the degrees of freedom transform as $N$-component vectors under a symmetry group. A well-known example is the $O(N)$ symmetric theory of $N$ scalar fields $\phi^a$ in $d$ dimensions with interaction
$g (\phi^a \phi^a)^2$ (for reviews see \cite{Wilson:1973jj,Moshe:2003xn}). 
It is exactly solvable in the large $N$ limit where $gN$ is held fixed, since summation over the necessary class of bubble diagrams is not hard to evaluate.
Another famous class of examples are models of interacting
$N\times N$ matrix fields, so that the number of degrees of freedom scales as $N^2$; here one can introduce single-trace interactions like $g \tr \phi^4$. A 
significant simplification occurs in the 't Hooft large $N$
limit where $g N$ is held fixed: the perturbative expansion is dominated by the planar diagrams \cite{'tHooft:1973jz}. While such planar matrix theories are exactly solvable in some special low-dimensional
cases \cite{Brezin:1977sv}, the problem does not appear to be solvable in general.  

\begin{figure}[h!]
                \centering
                \includegraphics[width=14cm]{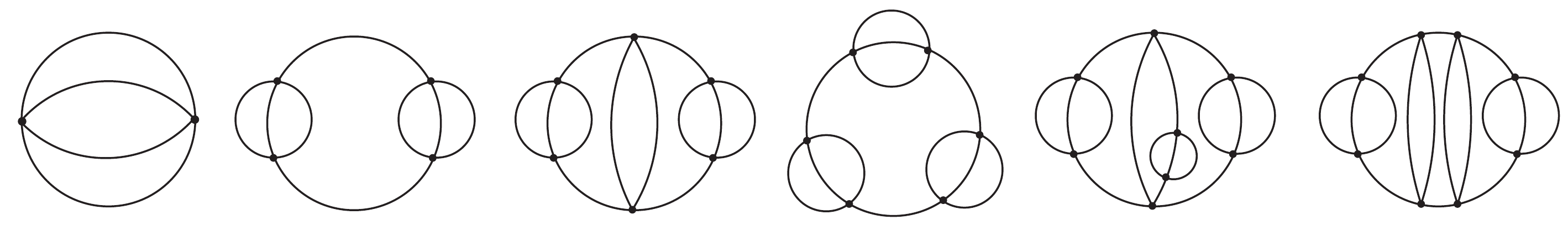}
\caption{Some melonic contributions to the free energy.}
                \label{MelonsEx}
\end{figure}

In view of these classic results, it is natural to study theories with rank-$m$ tensor degrees of freedom $\phi^{a_1 \ldots a_m}$, where each index takes $N$ values so that the
net number of degrees of freedom scales as $N^m$ \cite{Ambjorn:1990ge,Sasakura:1990fs,Gross:1991hx}. Since the complexity of taking the large $N$ limit increases from $m=1$ to $m=2$, one might expect that the tensor models
with $m>2$ are much more difficult than the matrix models.  
However, 
Gurau and collaborators \cite{Gurau:2009tw,Gurau:2011xp,Gurau:2011aq,Gurau:2011xq,Bonzom:2011zz,Bonzom:2012hw}
have discovered
that, by adjusting the interactions appropriately, it is possible to find models with $m>2$ where a large $N$ limit is solvable. The
perturbative expansion is then dominated by a special class
of ``melon diagrams" (for some examples with $m=3$ see figures \ref{MelonsEx}).

Gurau's original example \cite{Gurau:2009tw} was a so-called colored tensor model where complex fermionic tensors $\psi_A^{a_1 \ldots a_m}$ carry an additional label $A$ which takes $m+1$
possible values $0,1, \ldots m$. In the smallest non-trivial case $m=3$ this model has the interaction 
\begin{align}
g \psi_0^{a b c }\psi_1^{a d e}\psi_2^{f b e} \psi_3^{fdc} + {\rm c. c.}  \label{FermInter}
\end{align}
The label $A$ may be thought of as corresponding to the $4$ different vertices of a tetrahedron. Each pair of fields has one pair of indices in common, just as every pair of vertices
of a tetrahedron is connected by one edge. The interaction (\ref{FermInter}) has $U(N)^6$ symmetry, where each $U(N)$ corresponds to one of the
edges of the tetrahedron. Including the quadratic piece $\psi_A^{abc} \bar \psi_A^{abc}$ and integrating over  the fermionic tensors with interaction (\ref{FermInter}) 
generates a summation over a particular class of 3-dimensional intrinsic geometries made out of tetrahedra.
 Apart from this interpretation, this model is of much interest because it exhibits a novel type
of large $N$ limit, where the coupling constant is scaled so that $g^2 N^3$ is held constant, and the theory has $N^3$ degrees of freedom.\footnote{The $N^3$
scaling of the degrees of freedom is also found for 6-dimensional CFTs on $N$ coincident M5-branes \cite{Klebanov:1996un,Harvey:1998bx}. 
An interpretation of this scaling in terms of M2-branes with three holes attached to three different M5-branes, thus
giving rise to tri-fundamental matter, was proposed in section 5.2 of \cite{Klebanov:1996mh}. One may wonder if there is a precise connection between theories on
M5-branes and tensor models.} 
Thus, it is interesting to try generalizing
Gurau's construction\footnote{The random tensor models also have connections with
the ``holographic space-time" approach to quantum gravity \cite{Banks:2013fri,Banks:2016taq}.} from the $d=0$ tensor integral case to 
$d$-dimensional quantum theories. 
An important step in this direction was recently made by Witten \cite{Witten:2016iux}, who studied a quantum mechanical model of colored anti-commuting
tensors and observed
that its large $N$ limit is similar to that in the Sachdev-Ye-Kitaev (SYK) model \cite{Sachdev:1992fk,1999PhRvB..59.5341P, 2000PhRvL..85..840G,Kitaev:2015}.    

The quantum mechanical model introduced by Witten uses, in the $m=3$ case, real fermionic tensors $\psi_A^{abc}$ and possesses $O(N)^6$ symmetry.
The action of this model is
\begin{align}
S_{\rm Gurau-Witten}
 = \int d t \Big( \frac i 2 \psi_A^{abc}\partial_{t}\psi_A^{abc}+ g \psi_0^{a b c }\psi_1^{a d e}\psi_2^{f b e} \psi_3^{fdc} \Big)\,, \label{FermAct1}
\end{align}
It was shown \cite{Witten:2016iux,Gurau:2016lzk} that, in the large $N$ limit of this model only the ``melonic" Feynman graphs survive, just as in the SYK model. 
Very importantly, gauging the $O(N)^6$ symmetry gets rid of
the non-singlet states. This removes a crucial conceptual obstacle in the search for the gravity dual of this model, in the spirit of the AdS/CFT correspondence for gauge
theories \cite{Maldacena:1997re,Gubser:1998bc,Witten:1998qj}.  

In work subsequent to \cite{Gurau:2009tw} it was shown that the ``coloring" is not necessary for obtaining a large $N$ limit where the melon graphs dominate, 
and theories of just one complex bosonic tensor were shown to have this property \cite{Bonzom:2012hw,Tanasa:2011ur,Dartois:2013he,Tanasa:2015uhr}. 
More recently, a model of a single real
rank-$3$ tensor with $O(N)^3$ symmetry was studied by Carrozza and Tanasa and shown to possess a melonic large $N$ limit \cite{Carrozza:2015adg}.
We will study such a theory of one real rank-$m$ fermionic tensor with interaction $\psi^{m+1}$. For $m=3$ the interaction assumes explicit form
\begin{equation}
V_{\rm uncolored}=
\frac{1}{4}g \psi^{a_{1}b_{1}c_{1}}\psi^{a_{1}b_{2}c_{2}}\psi^{a_{2}b_{1}c_{2}}\psi^{a_{2}b_{2}c_{1}}
\label{uncoloredint}
\end{equation}
The three indices are distinguishable, and the theory has $O(N)^3$ symmetry under
\begin{align}
\psi^{abc} \to M_{1}^{aa'} M_{2}^{bb'}M_{3}^{cc'}\psi^{a'b'c'}, \quad M_{1},M_{2},M_{3} \in O(N)\,.
\end{align}
Thus, the real field $\psi^{abc}$ transforms in the tri-fundamental representation of $O(N)^3$. Such an uncolored fermionic model does not work in $d=0$ 
because the invariant quadratic term vanishes,
$\psi^{abc} \psi^{abc}=0$, but in $d=1$ there is a non-trivial model with the kinetic term $\frac i 2 \psi^{abc}\partial_{t}\psi^{abc}$.
We will also consider analogous bosonic models where the anti-commuting field in (\ref{uncoloredint}) is replaced by a commuting one, $\phi^{abc}$. 
Then in $d=0$ we may add the quadratic term  $\phi^{abc} \phi^{abc}$, while in $d>0$ the standard kinetic term $\frac{1}{2} \partial_{\mu}\phi^{abc}\partial^{\mu}\phi^{abc}$.
While the bosonic potential is generally not positive definite,\footnote{We thank E. Witten for pointing this out to us.}
the model may still be studied in perturbation theory. One may hope that, as in the matrix models, the restriction to leading large $N$ limit can formally stabilize the theory.
   
In section \ref{Theproof} we study the index structure of the expansion of the path integral in $g$ and
demonstrate that the large $N$ limit is dominated by the melon diagrams. \footnote{We constructed the argument before the existence of \cite{Carrozza:2015adg} was
pointed out to us, so it may provide an independent perspective on the $O(N)^3$ theories.} 
The argument, which applies to both the uncolored fermionic and bosonic models, 
contains a new ingredient compared to other models. In uncolored models with complex tensors, which were studied in \cite{Bonzom:2012hw},
each index loop necessarily passed through an even number of vertices, but in models with real tensors a loop can also pass through an odd number of vertices.
However, the diagrams dominant in the large $N$ limit do not contain any index loops that pass through 3 vertices. 

In section \ref{uncoloredqm} we show that the uncolored fermionic theory with interaction (\ref{uncoloredint}) is equivalent
 to the SYK model in the large $N$ limit. We comment on the
spectrum of operators in the gauged tensor models, pointing out that it appears to be vastly bigger than the ``single Regge trajectory" which has been studied in the
SYK model so far \cite{Polchinski:2016xgd,Maldacena:2016hyu,Jevicki:2016bwu,Gross:2016kjj}. 
In section \ref{compferm} we write down a $U(N)^2\times O(N)$ symmetric quantum mechanical model with a complex fermionic 3-tensor. 
We study the large $N$ limit of this model and derive the scaling dimensions of two-particle operators.
We argue that this model  is
related in the large $N$ limit to the generalization of SYK model which contains complex fermions \cite{Sachdev:2015efa,Fu:2016vas}.
In section \ref{unbos} we study the large $N$
limit of the uncolored bosonic model with $O(N)^3$ symmetry.
We derive the Schwinger-Dyson equation for the two-point function and obtain its solution which is consistent with conformal invariance. It gives scaling dimension
$\Delta_\phi=\frac {d} {4} + {\cal O}(1/N)$. We also derive the Schwinger-Dyson equation for the four-point function and find the scaling dimensions
of two-particle operators. In section \ref{unbos} we also mention models with only one $O(N)$ symmetry group and matter in the fully symmetric and traceless or anti-symmetric representations. In these cases we have checked the melonic dominance at large $N$
up to order $g^7$, but a general proof seems harder to construct. In section \ref{epsunbos} we check this result by a perturbative calculation in $4-\eps$ dimensions at large $N$.
In section \ref{disc} we discuss various possible extensions of our results, including supersymmetric models with quartic superpotentials for 3-tensor superfields.


\section{Melonic Dominance in the $O(N)^3$ Symmetric Theories}
\label{Theproof}

The arguments in this section, which are analogous to those in \cite{Carrozza:2015adg}, 
apply to the uncolored models with $O(N)^3$ symmetry, both in the fermionic and bosonic cases and for any $d$. 
We will ignore the coordinate dependence and just focus on the index structure.

 \begin{figure}[h!]
                \centering
                \includegraphics[width=3.5cm]{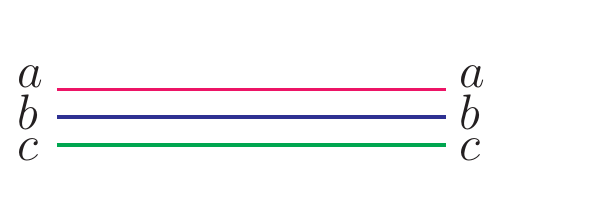}
                \caption{The resolved propagator $\langle \phi^{abc} \phi^{a'b'c'} \rangle = \delta^{aa'}\delta^{bb'}\delta^{cc'} $.}
                \label{Propagator}
\end{figure}

\begin{figure}[h!]
                \centering
                \includegraphics[width=6cm]{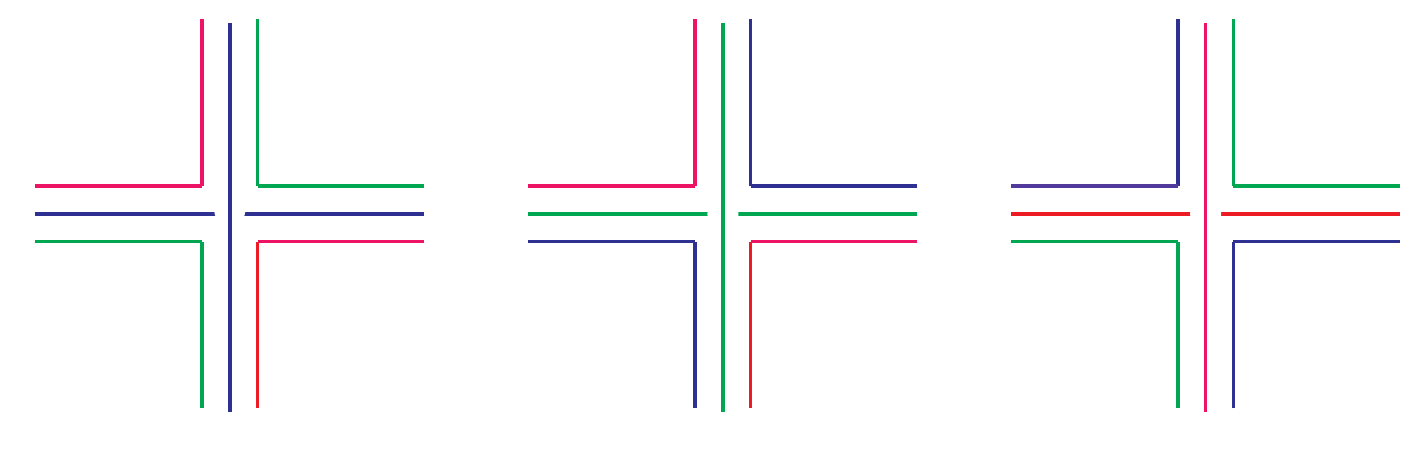}
                \caption{Three equivalent ways to represent the resolved vertex.}
                \label{vertex}
\end{figure}

\begin{figure}[h!]
                \centering
                \includegraphics[width=1.8cm]{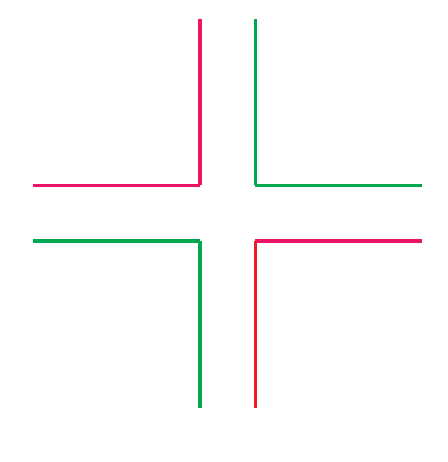}
                \caption{The standard matrix model vertex obtained after ``forgetting''  the middle lines.}
                \label{FatVertex}
\end{figure}

The propagator of the $\phi^{abc}$ field has the index structure depicted in figure \ref{Propagator}. 
The three colored wires (also called ``strands" in the earlier literature) represent propagation of the 
three indices of the $\phi^{abc}$ field. In spite of this coloring, the model is ``uncolored" in the standard terminology, since it contains only one tensor field.
The vertex has the index structure depicted in the figure \ref{vertex}. 
There are three equivalent ways to draw the vertex; for concreteness we will use the first way. "Forgetting" the middle lines we obtain the standard matrix model vertex as in figure \ref{FatVertex}. 

\begin{figure}[h!]
                \centering
                \includegraphics[width=12cm]{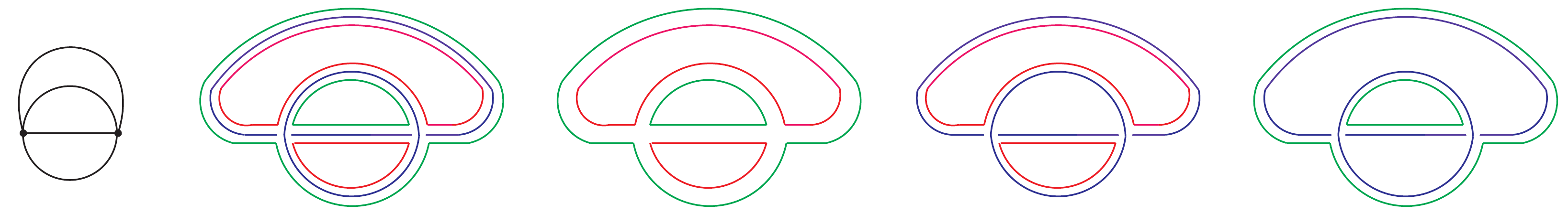}
                \caption{A melonic second-order diagram and all its fat subgraphs.}
                \label{ExampleDg2p3}
\end{figure}

\begin{figure}[h!]
                \centering
                \includegraphics[width=13cm]{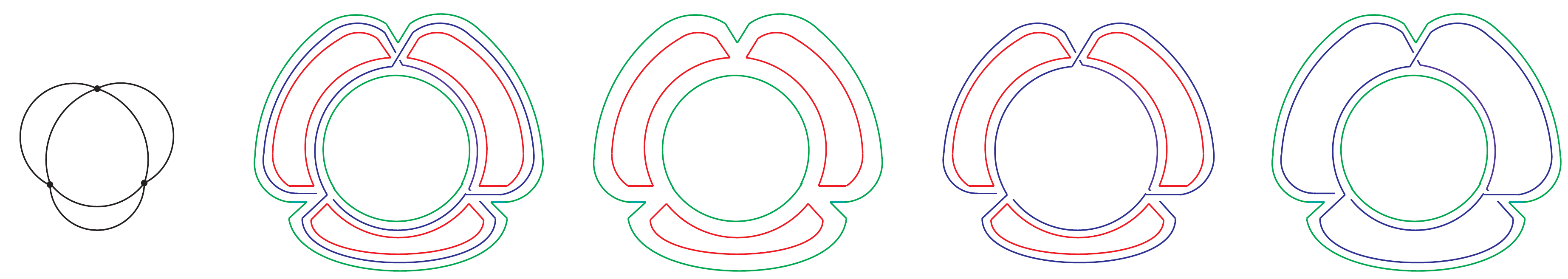}
                \caption{A non-melonic third-order diagram and all its fat subgraphs.}
                \label{ExampleDg3p3}
\end{figure}

Let us consider the vacuum Feynman diagrams. Examples of melonic and non-melonic diagrams with their resolved representations and fat (double-line) subgraphs  
are depicted in figures \ref{ExampleDg2p3} and  \ref{ExampleDg3p3}.

Each resolved Feynman diagram consists of  loops of three different colors and is proportional to $N^{f_{\rm total}}$, where $f_{\rm total}$ is the total number of index loops. 
Suppose we ``forget'' all wires of some particular color 
in our diagram, as in the pictures \ref{ExampleDg2p3} and  \ref{ExampleDg3p3}. Then we get a double-line fat graph (ribbon graph) of the kind one finds in
matrix models. One can count the number of all index loops $f$ in this fat graph using the Euler characteristic $\chi$
\begin{align}
f= \chi +e-v\,, \label{Euler}
\end{align}
where $e$ is the number of edges and $v$ is the number of vertices. In our theory we obviously have $e= 2v$, therefore $f= \chi+v$. 
We can forget red, blue or green wires, and in each case we get a fat graph made of the remaining two colors. If we forget, say, all red wires, then using the formula
(\ref{Euler}) we find $f_{bg}=\chi_{bg}+v$, where $f_{bg}=f_{b}+f_{g}$ is the number of blue and green loops and $\chi_{bg}$ is the Euler characteristic of this blue-green fat graph. Analogously we get $f_{rg}=\chi_{rg}+v$ and $f_{br}=\chi_{br}+v$. Adding up all these formulas we find 
\begin{align}
f_{bg}+f_{rg}+f_{br} = 2(f_{b}+f_{g}+f_{r})=\chi_{bg}+\chi_{br}+\chi_{rg}+3v\,.
\end{align}
Thus, the total number of loops is 
\begin{align}
f_{\rm total} = f_{b}+f_{g}+f_{r}= \frac{3v}{2}+3 -g_{bg}-g_{br}-g_{rg}\,, \label{fandgen}
\end{align}
where $g=1-\chi/2$ is the genus of a graph. Because $g\geqslant 0$ we obtain 
\begin{align} 
f_{\rm total} \leqslant 3+\frac{3v}{2}\,.
\end{align}
Now  the goal is to show that the equality $f_{\rm total} = 3+3v/2$ is satisfied  only for the melonic diagrams. We will call the graphs which 
satisfy $f_{\rm total}= 3+3v/2$ the maximal graphs. Thus we should argue that maximal graphs are necessarily melonic. 
We note that,  due to (\ref{fandgen}), each double-line fat subgraph of a maximal graph has genus zero.

Now let us classify all loops in our graph according to how many vertices they pass through (a loop can pass
the same vertex twice). Let us denote by $\mathcal{F}_{s}\geqslant 0$ the number of loops, which pass through $s$ vertices. 
For a maximal graph
\begin{align}
f_{\rm total}=\mathcal{F}_{2}+\mathcal{F}_{3}+\mathcal{F}_{4}+\mathcal{F}_{5}+\ldots = 3+ \frac{3v}{2}\ ,
\end{align}
where we set $\mathcal{F}_{1}=0$ because we assume that there are no tadpole diagrams.
Since each vertex must be passed $6$ times, we also get 
\begin{align} 
2\mathcal{F}_{2} +3\mathcal{F}_{3}+4\mathcal{F}_{4}+5\mathcal{F}_{5}+\dots = 6v\,.
\end{align}
Combining these two equations we find 
\begin{align} 
2\mathcal{F}_{2} +\mathcal{F}_{3} = 12+\mathcal{F}_{5}+2\mathcal{F}_{6}+\dots\,. \label{maineqforF}
\end{align}
Now our goal  is to show that $\mathcal{F}_{2} >0$ using this formula (in fact, $\mathcal{F}_{2} \geqslant 6$, but all we will need is that it is non-vanishing).

\begin{figure}[h!]%
    \centering
    \subfloat{{\includegraphics[width=3.5cm]{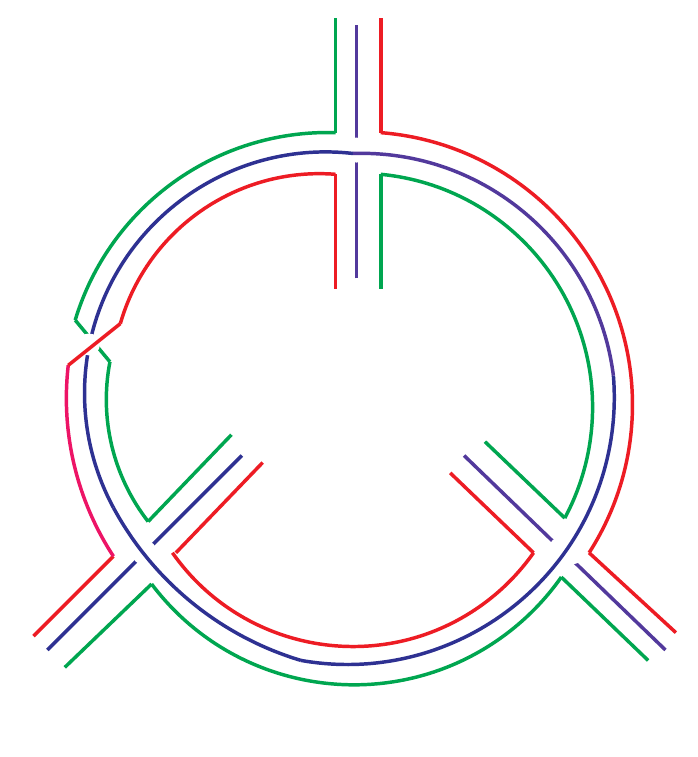} }}%
    \qquad
    \subfloat{{\includegraphics[width=3.5cm]{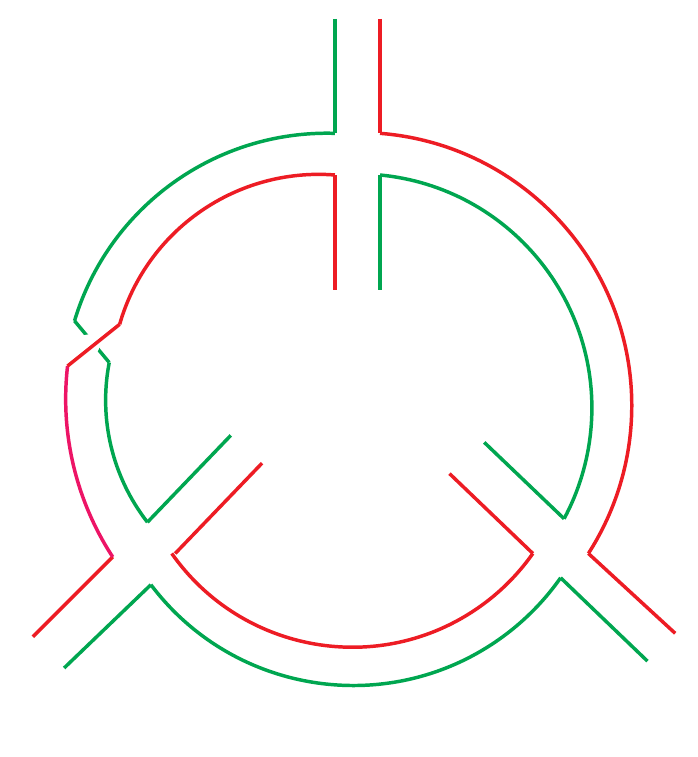} }}%
    \caption{a) Local part of a graph with a middle index loop passing  through 3 vertices. b) The same figure where the middle index has been ``forgotten." }
    \label{Twsited3vertLoop}%
\end{figure}

Let us first argue that a maximal graph must have $\mathcal{F}_{3}=0$.
To have $\mathcal{F}_{3}>0$ we need a closed index loop passing through 3 vertices. Without a loss of generality we can assume that this loop is formed by the middle lines in each vertex (blue lines). The only possibility with a closed loop of an internal (blue) index, which passes through three vertices, is shown in fig. \ref{Twsited3vertLoop} a). After "forgetting" the color of this loop we get a fat graph  in fig. \ref{Twsited3vertLoop} b), which is non-planar due a twisted propagator. 
So, a graph with $\mathcal{F}_{3}>0$ cannot be maximal. Thus, setting  $\mathcal{F}_{3}=0$ in (\ref{maineqforF}), 
we deduce that a maximal graph should have $\mathcal{F}_{2}>0$.

\begin{figure}[h!]%
    \centering
    \subfloat{{\includegraphics[width=3.5cm]{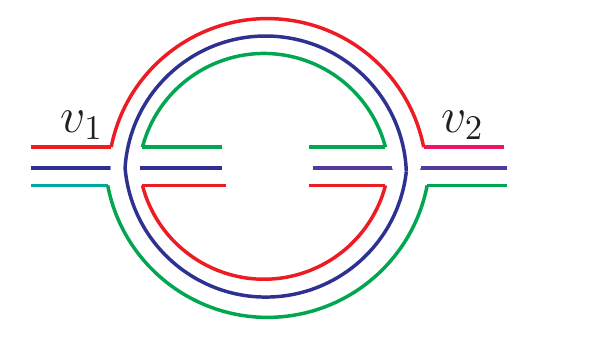} }}%
    \qquad
    \subfloat{{\includegraphics[width=3.5cm]{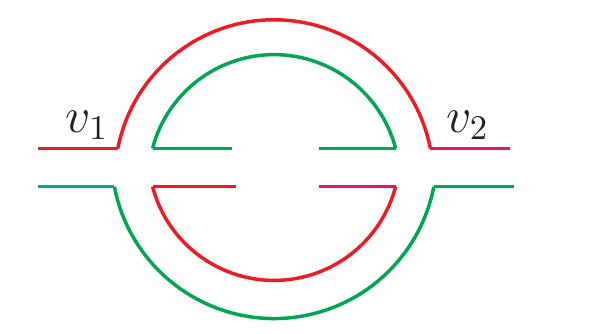} }}%
    \caption{a) Local part of a graph with a middle index loop passing  through two vertices $v_{1}$ and $v_{2}$. b) The same figure where the middle index has been ``forgotten." }
    \label{OnlyPossible2vert}%
\end{figure}

Finally, we need to show that the graphs with $\mathcal{F}_{2}>0$ are melonic. To do this
we will follow Proposition 3 in \cite{Bonzom:2011zz}.
Without a loss of generality we assume that the loop passing through $2$ vertices 
is formed by the middle lines in each vertex (blue lines). The only such possibility 
is shown in fig. \ref{OnlyPossible2vert} a). After "forgetting" the color of this loop we get a fat graph  in fig. \ref{OnlyPossible2vert} b).

\begin{figure}[h!]
                \centering
                \includegraphics[width=8cm]{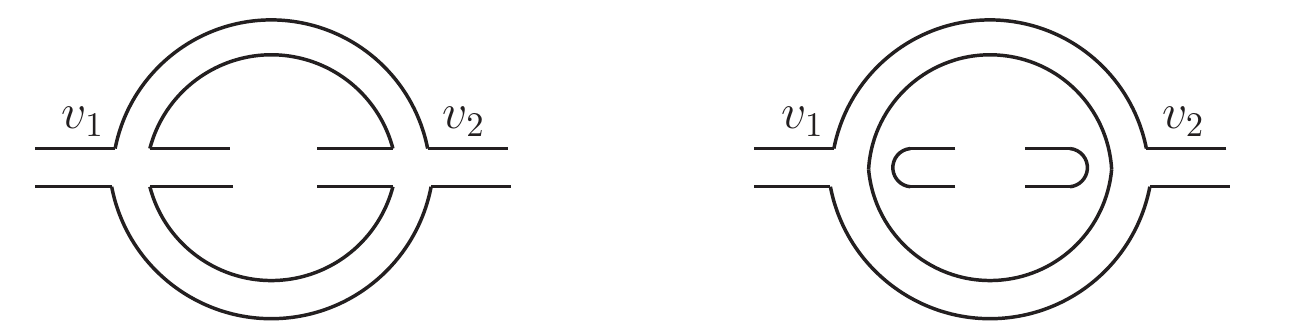}
                \caption{Cutting and sewing lines.}
                \label{2edgesCut}
\end{figure}

\begin{figure}[h!]
                \centering
                \includegraphics[width=3cm]{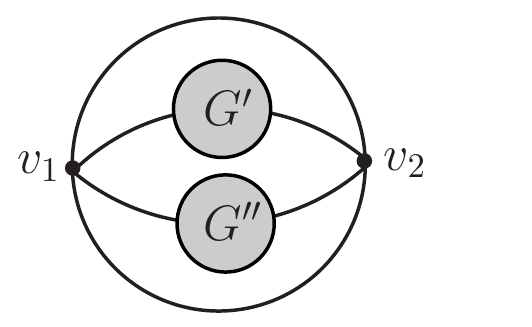}
                \caption{General structure of the maximal graph. }
                \label{GeneralPict1}
\end{figure}

\noindent  Now we uncolor the lines in our fat graph and cut  and sew two edges as in figure \ref{2edgesCut}.  
We cut two edges but did not change the number of loops; therefore, the Euler characteristic of the new graph is $\chi=4$. This is possible only if we separated our original graph into two  genus zero parts.  Therefore, our graph is two-particle reducible for the internal and external couples of lines. Thus, the  whole unresolved  graph  looks like figure \ref{GeneralPict1}.
Then, if graphs $G'$ and $G''$ are empty we get a second-order melon graph as in figure \ref{ExampleDg2p3}.
If they are not empty one can argue (see \cite{Bonzom:2011zz}) that they are also maximal graphs. So, we can recursively apply the same above argument  to them, implying 
that the complete diagram is melonic.

\section{Uncolored Quantum Mechanics and the SYK Model}
\label{uncoloredqm}

Using the interaction (\ref{uncoloredint}) we will now consider an ``uncolored" quantum mechanical model with real anti-commuting variables
$\psi^{abc}(t)$ and
the action 
\begin{align}
S = \int d t \Big( \frac i 2 \psi^{abc}\partial_{t}\psi^{abc}+ \frac{1}{4}g \psi^{a_{1}b_{1}c_{1}}\psi^{a_{1}b_{2}c_{2}}\psi^{a_{2}b_{1}c_{2}}\psi^{a_{2}b_{2}c_{1}}\Big)\, . \label{FermAct3}
\end{align}
It has $1/4$ of the degrees of freedom of the colored Gurau-Witten model (\ref{FermAct1}). We will argue that the uncolored model (\ref{FermAct3}) is equivalent to the SYK model
in the large $N$ limit.

We recall that $\psi^{abc}$ are the $N^3$ anticommuting fields and the indices, each of which runs from $1$ to $N$, are treated as distinguishable. 
The Fermi statistics implies 
\begin{align}
\psi^{a_{1}b_{1}c_{1}}\psi^{a_{1}b_{2}c_{2}}\psi^{a_{2}b_{1}c_{2}}\psi^{a_{2}b_{2}c_{1}}=
 -\psi^{a_{1}b_{2}c_{2}} \psi^{a_{1}b_{1}c_{1}} \psi^{a_{2}b_{1}c_{2}}\psi^{a_{2}b_{2}c_{1}} \,.
\end{align}
After relabeling 
$b_1 \leftrightarrow c_2$ and $b_2 \leftrightarrow c_1$ we get the relation 
\begin{align}
\psi^{a_{1}b_{1}c_{1}}\psi^{a_{1}b_{2}c_{2}}\psi^{a_{2}b_{1}c_{2}}\psi^{a_{2}b_{2}c_{1}}=
 -\psi^{a_{1}c_{1}b_{1}} \psi^{a_{1}c_{2}b_{2}} \psi^{a_{2}c_{2}b_{1}}\psi^{a_{2}c_{1}b_{2}} \ .
\end{align}
This demonstrates the vanishing of the interaction term in the $O(N)$ symmetric theory with a fully symmetric or fully anti-symmetric fermionic tensor.
Fortunately, in the theory with general 3-index fermionic tensors the interaction is non-trivial.

Let us return, therefore, to the theory (\ref{FermAct3}) with $O(N)^3$ symmetry, where the three indices are distinguishable.
The symmetry may be gauged by the replacement
\begin{equation}
\partial_t \psi^{abc}  \rightarrow (D_t \psi)^{abc} = \partial_t \psi^{abc} + A_1^{a a'} \psi^{a' b c} + A_2^{b b'} \psi^{a b' c} + A_3^{c c'} \psi^{a b c'}
\ ,
\end{equation}
where $A_i$ is the gauge field corresponding to the $i$-th $O(N)$ group. In $d=1$ the gauge fields are non-dynamical, and their only effect is to restrict the operators to be
gauge singlets. There is a sequence of such operators of the form 
\begin{equation}
O_2^n= \psi^{abc} (D_t^n \psi)^{abc}\ ,
\label{twoparticleops}
\end{equation}
where $n$ is odd. 
This set of operators is analogous to the ``single Regge trajectory" \cite{Polchinski:2016xgd,Maldacena:2016hyu,Gross:2016kjj} found in the Sachdev-Ye-Kitaev (SYK) model 
\cite{Sachdev:1992fk,1999PhRvB..59.5341P, 2000PhRvL..85..840G,Kitaev:2015}. 

\begin{figure}[h!]
                \centering
                \includegraphics[width=12cm]{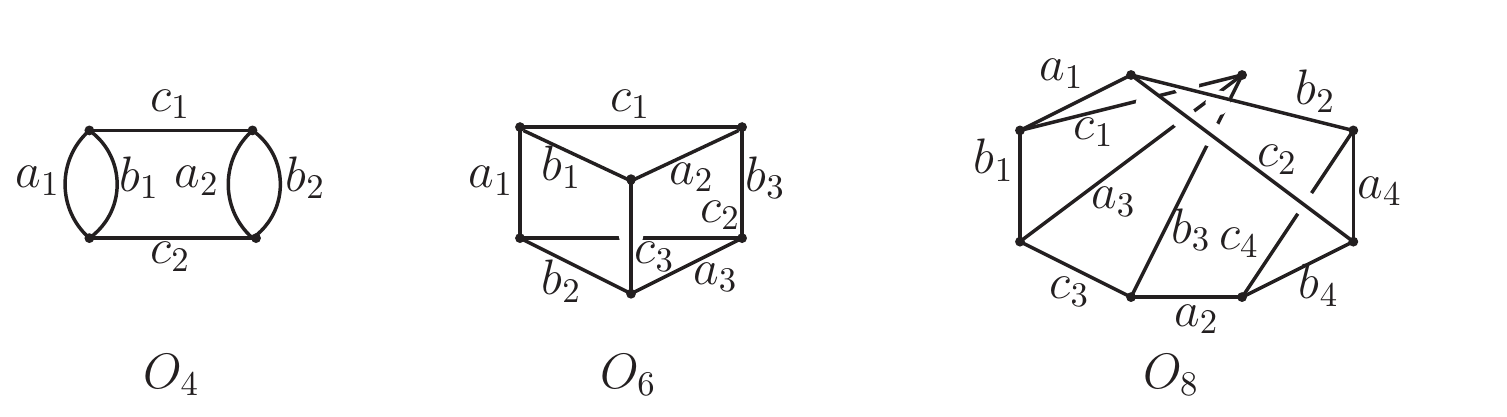}
                \caption{Graphical representation of different operators}
                \label{Alloprs}
\end{figure} 

We should note, however, that theory (\ref{FermAct3}) contains an abundance of additional ``single-trace"
$O(N)^3$ symmetric operators. A large class of them contains an even number of $\psi$ fields without derivatives and with all indices contracted. 
One of such $\psi^4$ operators is the interaction term in the action, which is related by the equation of motion to $\psi^{abc} (D_t \psi)^{abc}$.
Another type of $\psi^4$ operator is 
\begin{align}
O_4
=\psi^{a_{1}b_{1}c_{1}}\psi^{a_{1}b_{1}c_{2}}\psi^{a_{2}b_{2}c_{1}}\psi^{a_{2}b_{2}c_{2}}
\  ,
\label{pillow}
\end{align}
and there are similar operators where the second and third or the first and third indices have pairwise contractions 
(however, in the theory where the $O(N)^3$ symmetry is gauged such operators vanish because they are squares of the gauge symmetry generators).
Moving on to the higher operators we can try writing down the following $\psi^6$ operator:
\begin{align}
O_6
=\psi^{a_{1}b_{1}c_{1}}\psi^{a_{1}b_{2}c_{2}}\psi^{a_{2}b_{1}c_{3}}\psi^{a_{2}b_{3}c_{1}} \psi^{a_{3}b_{2}c_{3}}\psi^{a_{3}b_{3}c_{2}} 
\ .
\label{prism}
\end{align}
Due to the fermi statistics this operator actually vanishes, but an operator with $\psi$ fields replaced by scalars $\phi$ is present in the bosonic model that we study 
in section \ref{unbos}.
The following $\psi^8$ operator does not vanish in the fermionic model:
\begin{align}
O_8
=\psi^{a_{1}b_{1}c_{1}}\psi^{a_{1}b_{2}c_{2}}\psi^{a_{2}b_{3}c_{3}}\psi^{a_{2}b_{4}c_{4}} \psi^{a_{3}b_{1}c_{3}}\psi^{a_{3}b_{3}c_{1}} 
\psi^{a_{4}b_{2}c_{4}}\psi^{a_{4}b_{4}c_{2}}
\ .
\label{eightorder}
\end{align}
All such operators can be represented graphically with $\psi$-fields corresponding to vertices and index contractions to edges 
(see figure \ref{Alloprs}). These representations are similar to the Feynman diagrams in $\phi^{3}$ theory. A feature of the latter two operators is that each pair of $\psi$-fields has either one or no indices in common.
We expect to find an infinite class of operators of this type -- they should correspond to some number of tetrahedra glued together. Since there is no 
parametrically large dimension gap in the set of operator dimensions, the holographic dual of this theory should be highly curved.

\begin{figure}[h!]
                \centering
                \includegraphics[width=16cm]{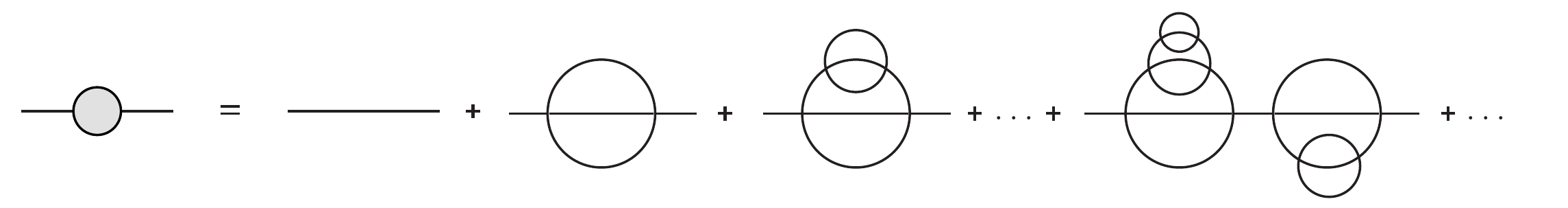}
                \caption{Diagrams contributing to the two point function in the leading  large $N$ order. The line with the gray circle represents the full two point function. Each simple line is the bare propagator. }
                \label{Fullprop}
\end{figure}

\begin{figure}[h!]
                \centering
                \includegraphics[width=9cm]{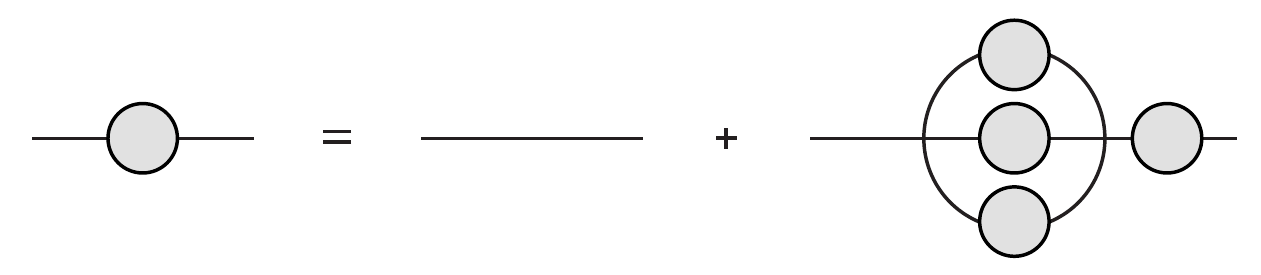}
                \caption{The graphical representation of the Schwinger-Dyson equation for the two-point function. }
                \label{SDeqpict}
\end{figure}

\begin{figure}[h!]
               \centering
              \includegraphics[width=16cm]{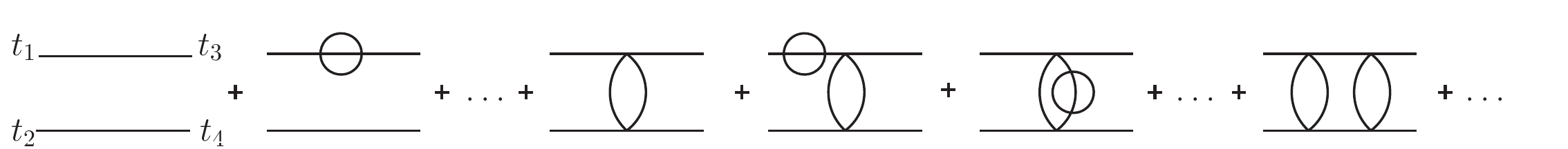}
              \caption{Ladder diagrams contributing to $\Gamma(t_{1},\dots, t_{4})$}
              \label{Ladder2}
\end{figure}

Let us study some of the diagrammatics of the uncolored quantum mechanics model (\ref{FermAct3}). We will study the ungauged model; the effect of the gauging may be
imposed later by restricting to the gauge invariant operators.
The bare propagator is 
\begin{align}
\langle T(\psi^{abc}(t)\psi^{a'b'c'}(0))\rangle_{0} =  \delta^{aa'}\delta^{bb'}\delta^{cc'}G_{0}(t)  =  \delta^{aa'}\delta^{bb'}\delta^{cc'} \frac{1}{2}\sgn (t)\,.
\end{align}
The full propagator in the large $N$ limit receives corrections from the melonic diagrams represented in figure \ref{Fullprop}. 
Resummation of all melonic diagrams  leads to the  Schwinger-Dyson equation for the two-point function
\begin{align}
G(t_{1}-t_{2}) = G_{0}(t_{1}-t_{2}) + g^{2}N^{3}\int dtdt' G_{0}(t_{1}-t)G(t-t')^{3} G(t'-t_{2})\,, \label{SD2pt}
\end{align}
represented graphically in figure \ref{SDeqpict}. This is the same equation as derived in \cite{Polchinski:2016xgd,Maldacena:2016hyu,Gross:2016kjj} for the large $N$ SYK model. 
The solution to (\ref{SD2pt}) in the IR limit is 
\begin{align}
G(t_{1}-t_{2})= -\Big(\frac{1}{4\pi g^{2}N^{3}}\Big)^{1/4}\ \frac{\sgn(t_{1}-t_{2})}{|t_{1}-t_{2}|^{1/2}}\,.
\label{twosolution}
\end{align}
To uncover the spectrum of the bilinear operators in the model, we need to study the 4-point function
$ \langle \psi^{a_1 b_1 c_1 } (t_1)  \psi^{a_1 b_1 c_1 } (t_2) \psi^{a_2 b_2 c_2 } (t_3)  \psi^{a_2 b_2 c_2 } (t_4) \rangle $. 
Its structure is again the same as in the large $N$ SYK model \cite{Maldacena:2016hyu, Polchinski:2016xgd}: 
\begin{align}
\langle \psi^{a_1 b_1 c_1 } (t_1)  \psi^{a_1 b_1 c_1 } (t_2) \psi^{a_2 b_2 c_2 } (t_3)  \psi^{a_2 b_2 c_2 } (t_4) \rangle =N^{6}G(t_{12})G(t_{34}) + \Gamma(t_{1},\dots ,t_{4})\,,
\end{align}
where $\Gamma(t_{1},\dots ,t_{4})$ is given by a series of ladder diagrams depicted in fig \ref{Ladder2}. 
 
\begin{figure}[h!]
               \centering
              \includegraphics[width=12cm]{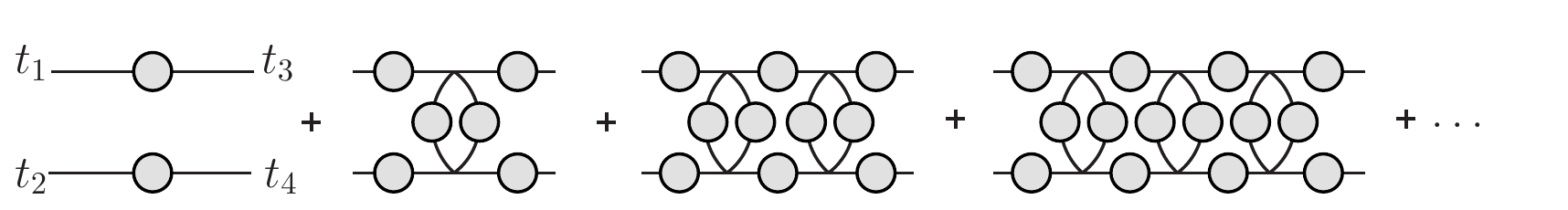}
              \caption{Ladder diagrams contributing to $\Gamma(t_{1},\dots, t_{4})$}
              \label{Ladder3}
\end{figure}

Resumming the diagrams in fig. \ref{Ladder2} one finds a contribution to $ \Gamma(t_{1},\dots ,t_{4})$ as a series of diagrams  in terms of the full  propagators, see fig. \ref{Ladder3}
 
\noindent If we denote by $\Gamma_{n}$ the ladder with $n$ rungs, so $\Gamma = \sum_{n} \Gamma_{n}$,  we have
\begin{align}
\Gamma_{0}(t_{1},\dots,t_{4})=N^{3}(-G(t_{13})G(t_{24})+G(t_{14})G(t_{23}))\,.
\end{align}
For the next coefficient one gets 
\begin{align}
\Gamma_{1}(t_{1},\dots,t_{4})=3g^{2}N^{6}\int dt dt'\big(G(t_{1}-t)G(t_{2}-t')G(t-t')^{2}G(t-t_{3})G(t-t_{4})-(t_{3} \leftrightarrow t_{4}) \big)\,,
\end{align}
and one can check further  that 
\begin{align}
\Gamma_{2}(t_{1},\dots,t_{4})=-3g^{2}N^{3}\int dt dt'\big(G(t_{1}-t)G(t_{2}-t')G(t-t')^{2}\Gamma_{1}(t,t',t_{3},t_{4})- (t_{3} \leftrightarrow t_{4})\big)\,.
\end{align}
So, in general, one gets exactly the same recursion relation as in  the SYK model 
\begin{align}
\Gamma_{n+1}(t_{1},\dots, t_{4}) = \int dt dt' K(t_{1},t_{2};t,t') \Gamma_{n}(t,t',t_{3},t_{4})\,,
\end{align}
where  the kernel is 
\begin{align}
K(t_{1},t_{2};t_{3},t_{4})  = -3 g^{2} N^{3}G(t_{13})G(t_{24})G(t_{34})^{2}\,.
\end{align}
In order to find the spectrum of the two-particle operators $O_{2}^n$, following \cite{Maldacena:2016hyu, Gross:2016kjj} one has to solve the integral eigenvalue equation
 \begin{align}
v(t_{0},t_{1},t_{2}) =g(h)\int dt_{3}dt_{4}K(t_{1},t_{2};t_{3},t_{4})v(t_{0},t_{3},t_{4})\,, \label{SDopeq}
\end{align}
where  
 \begin{align}
v(t_{0},t_{1},t_{2})=\langle O_{2}^n (t_{0})\psi^{abc}(t_{1})\psi^{abc}(t_{2})\rangle = \frac{\sgn(t_{1}-t_{2})}{|t_{0}-t_{1}|^{h}|t_{0}-t_{2}|^{h}|t_{1}-t_{2}|^{1/2-h}}\,,
\end{align}
is the conformal three-point function. Then the scaling dimensions are determined by the equation $g(h)=1$. To find $g(h)$ one can use $SL(2)$ invariance to take $t_{0}$ to infinity and just consider eigenfunctions of the form 
 \begin{align}
v(t_{1},t_{2})=\frac{\sgn(t_{1}-t_{2})}{|t_{1}-t_{2}|^{1/2-h}}\,. \label{antsymeigenf}
\end{align}
It is not hard to find $g(h)$ using two basic integrals 
 \begin{align}
&\int_{-\infty}^{+\infty}du \frac{\sgn(u-t_{1})\sgn(u-t_{2})}{|u-t_{1}|^{a}|u-t_{2}|^{b}}=l^{+}_{a,b}\frac{1}{|t_{1}-t_{2}|^{a+b-1}}\,,\notag \\
&\int_{-\infty}^{+\infty}du\frac{\sgn(u-t_{2})}{|u-t_{1}|^{a}|u-t_{2}|^{b}}=l^{-}_{a,b}\frac{\sgn(t_{1}-t_{2})}{|t_{1}-t_{2}|^{a+b-1}}\,, 
\notag \\
& l^{\pm}_{a,b} = \beta(1-a,a+b-1)\pm (\beta(1-b,a+b-1)-\beta(1-a,1-b))\ ,
\label{base1dint}
\end{align}
where 
$\beta(x,y)=\Gamma(x)\Gamma(y)/\Gamma(x+y)$
is the Euler beta function. Plugging (\ref{antsymeigenf}) into (\ref{SDopeq}) and using (\ref{base1dint}), we find \cite{Maldacena:2016hyu, Gross:2016kjj}
 \begin{align}
g(h)=-\frac{3}{4\pi}l^{+}_{\frac{3}{2}-h,\frac{1}{2}}l^{-}_{1-h,\frac{1}{2}}= -\frac{3}{2}\frac{\tan(\frac{\pi}{2}(h-\frac{1}{2}))}{h-1/2}\,.
\end{align}
The scaling dimensions are given by the solutions of $g(h)=1$. The first solution is exact, $h=2$; this is the important mode dual to gravity and
responsible for the quantum chaos in the model 
\cite{Almheiri:2014cka,Polchinski:2016xgd,Maldacena:2016hyu,Jevicki:2016bwu,Maldacena:2016upp,Engelsoy:2016xyb,Jensen:2016pah}.
The further solutions are $h\approx
3.77,\; 5.68,\; 7.63,\;9.60$ 
corresponding to the operators $\psi^{abc} (D_t^{n} \psi)^{abc}$ with $n=3,5,7,9$. In the limit of large $n$, $h_n\rightarrow n+\frac 1 2$. 
This is the expected limit $n+2 \Delta$, where $\Delta=\frac 1 4$ is the scaling dimension of the individual fermion.

\subsection{Models with a Complex Fermion}
\label{compferm}

Here we consider two quantum mechanical models of a complex $3$-tensor $\psi^{abc}$.
One of them is an uncolored version of the colored quantum mechanical model recently studied by Gurau \cite{Gurau:2016lzk}:
\begin{align}
S = \int d t \Big( i \bar \psi^{abc}\partial_{t}\psi^{abc}+ 
\frac{1}{4}g \psi^{a_{1}b_{1}c_{1}}\psi^{a_{1}b_{2}c_{2}}\psi^{a_{2}b_{1}c_{2}}\psi^{a_{2}b_{2}c_{1}}
+ \frac{1}{4} \bar g \bar \psi^{a_{1}b_{1}c_{1}}\bar \psi^{a_{1}b_{2}c_{2}}\bar \psi^{a_{2}b_{1}c_{2}}\bar \psi^{a_{2}b_{2}c_{1}}
\Big)\,, \label{FermAct5}
\end{align}
which again has $O(N)^3$ symmetry. Another possibility is the model
\begin{align}
S = \int d t \Big( i \bar \psi^{abc}\partial_{t}\psi^{abc}+ 
\frac{1}{2} g \psi^{a_{1}b_{1}c_{1}}\bar \psi^{a_{1}b_{2}c_{2}}\psi^{a_{2}b_{1}c_{2}} \bar \psi^{a_{2}b_{2}c_{1}}
\Big)\,, \label{FermAct7}
\end{align}
where the symmetry is enhanced to 
$U(N)\times O(N) \times U(N)$ because the transformations on the first and the third indices of the tensor are allowed to be $U(N)$.
Models of this type have been studied in $d=0$ \cite{Tanasa:2011ur,Dartois:2013he,Tanasa:2015uhr}.  
Gauging this symmetry in the quantum mechanical model restricts the operators to the singlet sector, allowing for the existence of a gravity dual.
The gauge invariant two-particle operators have the form 
\begin{align}
{\cal O}_2^{n}= \bar \psi^{abc} (D_t^{n} \psi)^{abc}\, \qquad n=0,1, \ldots \ ,
\end{align}
which includes $\bar \psi^{abc} \psi^{abc}$. There is also a variety of operators made out of the higher powers of the fermionic fields similarly to the
operators  (\ref{pillow}), (\ref{prism}), (\ref{eightorder}) in the $O(N)^3$ symmetric model of real fermions.
As established in  \cite{Tanasa:2011ur,Dartois:2013he,Tanasa:2015uhr},
the large $N$ limit of the complex uncolored model (\ref{FermAct7}) is once again given by the melon diagrams
(the arguments are easier than in \ref{Theproof} since each index loop passes through an even number of vertices). 
The  large $N$ limit of this model appears to be related
to the variant of SYK model where the real fermions are replaced by the complex ones \cite{Sachdev:2015efa,Fu:2016vas}.   

Let us briefly discuss summing over melonic graphs 
in the model (\ref{FermAct7}) at large $N$. 
The two-point function has the structure
\begin{align}
\langle T(\bar \psi^{abc}(t)\psi^{a'b'c'}(0))\rangle  =  \delta^{aa'}\delta^{bb'}\delta^{cc'}G (t) ,
\end{align}
and $G(t)=- G(-t)$. We find the same Schwinger-Dyson equation as (\ref{SD2pt}); its solution is again (\ref{twosolution}) 
indicating that the fermion scaling dimension is $\Delta=1/4$.
Now we need to study the 4-point function
$ \langle \bar \psi^{a_1 b_1 c_1 } (t_1)  \psi^{a_1 b_1 c_1 } (t_2) \bar \psi^{a_2 b_2 c_2 } (t_3)  \psi^{a_2 b_2 c_2 } (t_4) \rangle $. It leads to the same
integral eigenvalue equation as (\ref{SDopeq}), but with kernel \footnote{We are grateful to J. Maldacena
and D. Stanford for a very useful discussion about this, which helped us correct the normalization of (\ref{gsym}).}
\begin{align}
K(t_{1},t_{2};t_{3},t_{4})  =- g^{2} N^{3} \big ( 2 G(t_{13})G(t_{24})G(t_{34})^{2} -  G(t_{14})G(t_{23})G(t_{34})^{2}\big )\ .
 \end{align}
Now it is possible to have
not only the antisymmetric eigenfunctions as 
in (\ref{antsymeigenf}), but also the symmetric ones
 \begin{align}
v(t_{1},t_{2})=\frac{1}{|t_{1}-t_{2}|^{1/2-h}}\, . \label{antsymeigens}
\end{align} 
This can be justified by noticing that the three point function now is  $\langle {\cal O}^{n}_{2}(t_{0})\psi^{abc}(t_{1})\bar{\psi}^{abc}(t_{2})\rangle$.
We see that for odd $n$ it is 
antisymmetric under $t_{1}\leftrightarrow t_{2}$, while for even $n$ it is symmetric.

Substituting ansatz (\ref{antsymeigens}) into the integral equation, and using the integrals (\ref{base1dint}), we find
 \begin{align}
g_{\rm sym}(h)=-\frac{1}{4\pi}l^{-}_{\frac{3}{2}-h,\frac{1}{2}}l^{+}_{1-h,\frac{1}{2}}= -\frac{1}{2}\frac{\tan(\frac{\pi}{2}(h+\frac{1}{2}))}{h-1/2}\,.
\label{gsym}
\end{align}
The scaling dimensions of the operators ${\cal O}_2^n$ with even $n$ are given by the solutions of
 $g_{\rm sym} (h)=1$. The first eigenvalue is $h=1$, corresponding to the conserved $U(1)$ charge $\bar \psi^{abc} \psi^{abc}$.
The additional values are $h\approx 2.65,\; 4.58,\; 6.55,\; 8.54$ corresponding to the operators with $n=2,4,6,8$ respectively. 
For large $n$ the scaling dimensions approach $n+ \frac 1 2 $ as expected. The numerical results are in good agreement with the asymptotic formula \cite{Maldacena:2016hyu}
\begin{align}
h_n= n+ \frac 1 2 + \frac {1} {\pi n} + {\cal O}(n^{-3})
\end{align}
for $n>2$. 
For  ${\cal O}_2^n$ with odd $n$ the spectrum is the same as for the two-particle operators
(\ref{twoparticleops}) in the model with
$O(N)^3$ symmetry.

\section{Uncolored bosonic tensors}
\label{unbos}

In this section we consider the $d$-dimensional field theory of a real commuting tensor field
$\phi^{abc}(x)$ with distinguishable indices $a,b,c=1,\dots,N$:
\begin{align}
S = \int d^{d}x \Big( \frac{1}{2} \partial_{\mu}\phi^{abc}\partial^{\mu}\phi^{abc} + \frac{1}{4} g 
\phi^{a_{1}b_{1}c_{1}}\phi^{a_{1}b_{2}c_{2}}\phi^{a_{2}b_{1}c_{2}}\phi^{a_{2}b_{2}c_{1}}\Big)\,, \label{Tenac3}
\end{align}
This is the bosonic analogue of the uncolored fermionic theory with interaction (\ref{uncoloredint}); it again has $O(N)^3$ symmetry.
A feature of this theory is that the interaction potential is not bounded from below for $N>2$. For $N=2$ the potential may be written as a sum of squares,
but for $N>2$ we have explicitly checked that there is a negative direction. Nevertheless, we may consider formal perturbation theory in $g$.

The argument in section \ref{Theproof} that the melonic diagrams dominate in the large $N$ limit applies both to the fermionic and bosonic version of the theory
in any dimension $d$. We may therefore resum all such diagrams 
and derive the exact Schwinger-Dyson equation similar to that in \cite{Gurau:2011xp, Gurau:2011xq, Gurau:2011aq, Gurau:2009tw, Gurau:2012ix}. 
Let us explain this using a simple example of the two-point function in the theory (\ref{Tenac3}).

We have for the bare propagator   
\begin{align}
\langle \phi^{abc}(p)\phi^{a'b'c'}(-p) \rangle_{0}  = G_{0}(p)\delta^{aa'}\delta^{bb'}\delta^{cc'} =\frac{1}{p^{2}}\delta^{aa'}\delta^{bb'}\delta^{cc'}  \,.
\end{align}
In the large $N$ limit one gets the same  Schwinger-Dyson equation for the full two-point function as in  (\ref{SD2pt}), which we can write in the momentum space as
\begin{align}
G(p) = G_{0}(p) + \lambda^{2}G_{0}(p)\Sigma(p) G(p)\,, \label{SD1bos}
\end{align}
where we introduced the coupling $\lambda= g N^{3/2}$, which is held fixed in the large $N$ limit and 
 \begin{align}
 \Sigma(p) = \int \frac{d^{d}k d^{d}q}{(2\pi)^{2d}} G(q)G(k)G(p+q+k)\,.\label{SD2}
\end{align}
One can rewrite (\ref{SD1bos}) as 
\begin{align}
G^{-1}(p) = G^{-1}_{0}(p) - \lambda^{2}\Sigma(p) \,. \label{SD3}
\end{align}
In the IR limit we can neglect the bare term $G_{0}(p)$ and get 
 \begin{align}
G^{-1}(p) = - \lambda^{2}\int \frac{d^{d}k d^{d}q}{(2\pi)^{2d}} G(q)G(k)G(p+q+k)\,. \label{SDeq1}
\end{align}
Using the integral
\begin{align}
\int \frac{d^{d}k}{(2\pi)^{d}}\frac{1}{k^{2\alpha}(k+p)^{2\beta}} = \frac{1}{(4\pi)^{d/2}}\frac{\Gamma(d/2-\alpha)\Gamma(d/2-\beta)\Gamma(\alpha+\beta-d/2)}{\Gamma(\alpha)\Gamma(\beta)\Gamma(d-\alpha-\beta)}\frac{1}{(p^{2})^{\alpha+\beta-d/2}} \label{mainint}
\end{align}
it is not difficult to show that the solution to the equation (\ref{SDeq1}) is
\begin{align}
G(p) = \lambda^{-1/2}\Big(\frac{(4\pi)^{d}d \Gamma(\frac{3d}{4})}{4\Gamma(1-\frac{d}{4})}\Big)^{1/4}\frac{1}{(p^{2})^{\frac{d}{4}}}\,. \label{2ptansw}
\end{align}
Alternatively, one can work in the coordinate representation and use the Fourier transform  
\begin{align}
\int d^{d}x \frac{e^{ikx}}{(x^{2})^{\alpha}} =\frac{\pi^{d/2}\Gamma(d/2-\alpha)}{2^{2\alpha-d}\Gamma(\alpha)} \frac{1}{(k^{2})^{d/2-\alpha}}
\end{align}
 to find the solution of the equation $G^{-1}(x)=-\lambda^{2}G^{3}(x)$:
\begin{align}
G(x)= \lambda^{-1/2} \Big(\frac{d \Gamma(\frac{3d}{4})}{4\pi^{d}\Gamma(1-\frac{d}{4})}\Big)^{1/4}\frac{1}{(x^{2})^{\frac{d}{4}}}\,.
\end{align}

If one works with the cutoff regularization, then the UV divergence, which arises in the integrals can be absorbed into mass renormalization.
Remarkably, the Schwinger-Dyson equation (\ref{SDeq1}) was originally studied in 1964, and its $d=3$ solution (\ref{2ptansw}) was found \cite{Patash:1964sp}.\footnote{
We thank A. Polyakov for pointing this out to us.} 

To find the spectrum of two-particle operators, we must solve for the eigenvalues $g_{\rm bos}(h)$ and eigenvectors $v_{h}$ of the kernel \cite{Polchinski:2016xgd,Maldacena:2016hyu, Gross:2016kjj} 
\begin{align}
\int d^{d}x_{3}d^{d}x_{4}K(x_{1},x_{2};x_{3},x_{4})v_{h}(x_{3},x_{4})= g_{\rm bos}(h) v_{h}(x_{1},x_{2})\, ,
\end{align}
where  the kernel is\footnote{We thank S. Giombi for correcting the sign error in the kernel that was present in an earlier version of this paper.}
\begin{align}
K(x_{1},x_{2};x_{3},x_{4})  = 3 \lambda^{2}G(x_{13})G(x_{24})G(x_{34})^{2}\,.
\end{align}
It is not hard to check using the integral (\ref{mainint}) that there is a subset of spin-zero eigenvectors 
\begin{align}
v_{h}(x_{1},x_{2}) = \frac{1}{[(x_{1}-x_{2})^{2}]^{\frac{1}{2}(\frac{d}{2}-h)}}\,,
\end{align}
and the corresponding eigenvalues are
\begin{align}
g_{\rm bos}(h)=-\frac{3 \Gamma \left(\frac{3 d}{4}\right) \Gamma \left(\frac{d}{4}-\frac{h}{2}\right) \Gamma \left(\frac{h}{2}-\frac{d}{4}\right)}{\Gamma \left(-\frac{d}{4}\right) \Gamma \left(\frac{3 d}{4}-\frac{h}{2}\right) \Gamma \left(\frac{d}{4}+\frac{h}{2}\right)}\,.
\end{align}
The scaling dimensions $h$ of spin zero two-particle operators are then determined by solving $g_{\rm bos} (h)=1$. 
In $d=1$ the smallest positive solution is $h=2$, suggesting the existence of a gravity dual.
We plan to study a more complete set of scaling dimensions in general $d$ in future work.

We have also studied the effect of replacing in (\ref{Tenac3}) the general $3$-index tensor by the completely symmetric and traceless tensor field $\phi^{abc}$.
Such a theory would have a single $O(N)$ symmetry under 
\begin{align}
\phi^{abc} \to M^{aa'}M^{bb'}M^{cc'}\phi^{a'b'c'}\,, \quad M \in O(N) \,.
\end{align}
The corresponding fermionic model would be trivial due to the vanishing of the interaction, but the bosonic model is non-trivial. 
The key question is whether the leading contribution in $N$ comes from the melonic diagrams only.
We have checked all the vacuum diagrams up to order $g^7$ and did not find any violation of this rule (see p.257 in \cite{Kleinert:2001ax} for pictures of all vacuum diagrams up to order $g^7$), but we have not constructed a proof to all orders yet.
If the $O(N)$ symmetric theory of a symmetric traceless tensor is indeed melonic, then the derivation of the Schwinger-Dyson equation and its solution goes through
just as for the $O(N)^3$ symmetric theory of a general tensor. 

Finally, we may wonder if for theories with a single $O(N)$ group we may consider matter in other irreducible representations, such as completely antisymmetric
or mixed symmetry. Such theories also appear to be melonic at low orders in perturbation theory, but a general proof to all orders has not been constructed.

\subsection{$\eps$-expansion of a scaling dimension}
\label{epsunbos}

Let us consider the uncolored  bosonic model in $d=4-\eps$ dimension. Introducing renormalized fields and 
couplings and using an auxiliary scale $\mu$, we can compactly write the action (\ref{Tenac3}) in the form 
\begin{align}
S = \int d^{d}x \Big(  \frac{1}{2}(\partial_{\mu}\vec{\phi})^{2}+\frac{1}{4}  \mu^{\eps} g 
\vec{\phi}^{4}+\frac{1}{2}\delta_{\phi}(\partial_{\mu}\vec{\phi})^{2}+\frac{1}{4}\mu^{\eps}\delta_{g} \vec{\phi}^{4}\Big)\,, \label{Tenaceps}
\end{align}
where $\vec{\phi}= \phi^{abc}$ and $\vec{\phi}^{4}\equiv \phi^{a_{1}b_{1}c_{1}}\phi^{a_{1}b_{2}c_{2}}\phi^{a_{2}b_{1}c_{2}}\phi^{a_{2}b_{2}c_{1}}$. 
The latter is not the only quartic term allowed by the $O(N)^3$ symmetry. To renormalize the theory at finite $N$ we need to include two additional operators:
the double-trace operator $ O_{\rm double-trace}= (\phi^{abc} \phi^{abc})^2$ and the ``pillow operator" \cite{Carrozza:2015adg}
\begin{align}
O_{\rm pillow}=\phi^{a_{1}b_{1}c_{1}}\phi^{a_{1}b_{1}c_{2}}\phi^{a_{2}b_{2}c_{1}}\phi^{a_{2}b_{2}c_{2}}+
\phi^{a_{1}b_{1}c_{1}}\phi^{a_{2}b_{1}c_{1}}\phi^{a_{1}b_{2}c_{2}}\phi^{a_{2}b_{2}c_{2}}+
\phi^{a_{1}b_{1}c_{1}}\phi^{a_{1}b_{2}c_{1}}\phi^{a_{2}b_{1}c_{2}}\phi^{a_{2}b_{2}c_{2}}
\ .\end{align}
In this section we carry out just the leading large $N$ analysis of operator dimension $\Delta_\phi$, where we believe the effects of these additional operators may be
ignored.

\begin{figure}[h!]
               \centering
              \includegraphics[width=2.5cm]{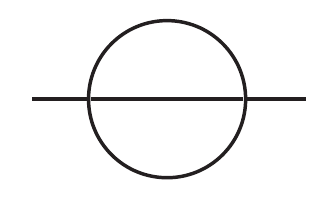}
              \caption{Two-loop diagram contributing to the anomalous dimension of the $\phi$ field.}
              \label{D11}
\end{figure}

The bare coupling is related to the renormalized one as 
\begin{align}
&g_{0}=\mu^{\eps}Z_{g}Z_{\phi}^{-2} g\,,
\end{align}
where $Z_{g}=1+\delta_{g}/g $ and $Z_{\phi}= 1+\delta_{\phi}$.
The bare propagator is 
\begin{align}
\langle \phi^{abc}(p)\phi^{a'b'c'}(-p)\rangle_{0} =  \delta^{aa'}\delta^{bb'}\delta^{cc'}\frac{1}{p^{2}}\,.
\end{align}
To compute the anomalous dimension of $\phi$ we consider the diagram in fig. \ref{D11}.  

\noindent We find at $d=4-\eps$ in the large $N$ limit
 \begin{align}
\delta_{\phi} =-\frac{g^{2}N^{3}}{2 (4 \pi )^4 \epsilon }\,, \quad \gamma_{\phi} = \frac{g^{2}N^{3}}{2 (4 \pi )^4  }\,.
\end{align}
On the other hand for the 4-point function  all  the one- and two-loop diagrams are subleading in $N$. Therefore, the beta-function is defined by the counter-term $\delta_{\phi}$
 \begin{align}
g_{0} =\mu^{\eps}(g-2\delta_{\phi}g ) =\mu^{\eps}\Big(g+\frac{g^{3}N^{3}}{ (4 \pi )^4 \epsilon } +\dots\Big)\,.
\end{align}
In the large $N$ limit where $g^2 N^3$ is held fixed, the beta-function is 
 \begin{align}
\beta_{g} = -\epsilon g + \frac{2 N^{3} g^{3}}{(4\pi)^{4}}+\ldots\,.
\end{align}
The theory in $4-\epsilon$ dimensions has a weakly coupled IR fixed point at 
 \begin{align}
g_{*}^{2} = \frac{(4\pi)^{4} \epsilon}{2N^{3}} \, ,
\end{align}
where the anomalous dimension at the critical point is 
 \begin{align}
 \gamma_{\phi} = \frac{\eps}{4}\,.
\end{align}
This leads to 
\begin{align}
\Delta_\phi= \frac{d-2} {2}+ \gamma_{\phi} = 1- \frac{\eps}{4} + {\cal O} (\eps^2) \, ,
\end{align}
which agrees with the large $N$ scaling dimension $\frac d 4$ obtained in (\ref{2ptansw}). It would be interesting to extend this perturbative analysis of the melonic
$\phi^4$ theory to higher orders in $\epsilon$ and to also include the $1/N$ corrections.

\section{Discussion}
\label{disc}

The existence of quantum mechanical models without disorder which admit a novel large $N$ limit dominated by the melonic graphs, such as the colored models explored
in \cite{Witten:2016iux,Gurau:2016lzk} and the uncolored models in section \ref{uncoloredqm}, opens new avenues for further research.
We have shown that various aspects of the $O(N)^3$ symmetric uncolored tensor model (\ref{FermAct3}) 
agree in the large $N$ limit with the SYK model \cite{Sachdev:1992fk,1999PhRvB..59.5341P, 2000PhRvL..85..840G,Kitaev:2015}.
Our uncolored tensor model is similar to the colored Gurau-Witten model (\ref{FermAct1}).
In particular, both models possess the same universal ``Regge trajectory" of two-particle operators as has been uncovered in the SYK model \cite{Polchinski:2016xgd,Maldacena:2016hyu,Gross:2016kjj}. In the uncolored model these are operators
(\ref{twoparticleops}), while 
 in the Gurau-Witten model they are $\psi_A^{abc} (D_t^{n} \psi_A)^{abc}$.
It would be interesting to carry out a more detailed comparison between the colored and uncolored models. 
As we have discussed, the uncolored model has a tower
of gauge invariant operators $\psi^n$. The same is true for the Gurau-Witten model; for example, at eighth order we find the operator
\begin{equation} 
O_8= \psi_{0}^{a_{1}b_{1}c_{1}} \psi_{1}^{a_{1}d_{1}e_{1}} \psi_{0}^{a_{2}b_{1}c_{2}} \psi_{1}^{a_{2}d_{1}e_{1}} \psi_{0}^{a_{3}b_{2}c_{1}} \psi_{1}^{a_{3}d_{2}e_{2}} \psi_{2}^{f_{1}b_{2}e_{2}} \psi_{3}^{f_{1}d_{2}c_{2}}  \ ,
\end{equation}
and similar operators with other choices of colors. The details of the operator spectra are not the same, however: due to the extra color label
the Gurau-Witten model contains more gauge invariant operators than our uncolored model.

In the uncolored model, in addition to the quartic operator in the action (\ref{FermAct3}) there are quartic operators of the form
(\ref{pillow}) (however, in the theory where the $O(N)^3$ symmetry is gauged such operators vanish). 
These kinds of operators are also present in the colored model, such as 
$\psi_0^{abc} \psi_0^{fbc} \psi_1^{ade} \psi_1^{fde}$ and analogous operators with other choices of the color labels.
In \cite{Carrozza:2015adg} such ``pillow operators" were included in the action and shown not to destroy the melonic dominance in the large $N$ limit.
Thus, imposing the $O(N)^3$ invariance appears to produce a class of quartic quantum mechanical models rather than a unique model. 
This is reminiscent of the fact that, in the $SU(N)$ symmetric quantum mechanics of a hermitian matrix $\Phi$ with potential $\tr \Phi^4$, one can add a double-trace term 
$(\tr \Phi^2)^2$, which can modify the free energy even in the leading large $N$ limit 
\cite{Das:1989fq, Sugino:1994zr, Gubser:1994yb }. 
The operators (\ref{pillow}) in the tensor model seem analogous to the double-trace operators in the matrix model, and their effect needs to be studied carefully.  

Some of the recent interest in the SYK model is related to the fact that it exhibits quantum chaos 
\cite{Polchinski:2016xgd,Maldacena:2016hyu,Jevicki:2016bwu,Maldacena:2016upp,Engelsoy:2016xyb,Jensen:2016pah}.
This was investigated numerically at finite $N$, providing further insights \cite{Fu:2016yrv,Garcia-Garcia:2016mno,Cotler:2016fpe}. 
Since at large $N$ the tensor quantum mechanical models become equivalent to the SYK model, one would expect them
to be chaotic as well, at least for sufficiently large $N$. A numerical investigation of the energy levels and thermal partition functions in the finite $N$
melonic quantum mechanical models should
be possible. The procedure would be quite different from that in \cite{Fu:2016yrv,Cotler:2016fpe} 
because there is no averaging over disorder. This may facilitate such a numerical study in the context of tensor quantum mechanics. 
 
It is also very interesting to ask if there exist quantum field theories
in dimensions above 1, which possess such a melonic large $N$ expansion. In section \ref{unbos} we began to study a bosonic $\phi^4$ tensor model which is renormalizable in $d=4$ and by power counting
may flow to a CFT in dimensions below 4. However, such a theory has the potential unbounded from below for $N>2$, so it does not appear to be stable at finite $N$. 

Another interesting possibility is to consider a supersymmetric theory with rank-$3$ tensor superfields $\Phi^{abc}$ and superpotential
\begin{align}
W= \frac{1} {4} g
\Phi^{a_{1}b_{1}c_{1}}\Phi^{a_{1}b_{2}c_{2}}\Phi^{a_{2}b_{1}c_{2}}\Phi^{a_{2}b_{2}c_{1}}
\ .
\label{superpot}
\end{align}
A simple setting for such a superspace approach is the supersymmetric quantum mechanics 
\cite{Witten:1981nf,Witten:1982df}, where (see, for example, \cite{Takeda:1985dg})
\begin{align}
\Phi^{a_{1}b_{1}c_{1}}(t, \theta, \bar \theta)= \phi^{a_{1}b_{1}c_{1}}(t) + i \theta \psi^{a_{1}b_{1}c_{1}}(t)-  i \bar \psi^{a_{1}b_{1}c_{1}}(t) \bar\theta 
+\bar \theta \theta F^{a_{1}b_{1}c_{1}}(t)\ ,
\end{align}
so that the degrees of freedom consist of a real bosonic 3-tensor and a complex fermionic one. The action may then be written as
\begin{align}
S= \int dt d\theta d\bar \theta \left (\frac 1 2 |D_\theta \Phi^{abc}|^2 + W \right ) \ ,
\end{align} 
and in terms of components it contains the two-boson two-fermion terms, such as 
\begin{equation}
\int dt (g \bar \psi^{a_{1}b_{1}c_{1}}\phi^{a_{1}b_{2}c_{2}}\psi^{a_{2}b_{1}c_{2}}\phi^{a_{2}b_{2}c_{1}} + {\rm c.\ c.} )
\ .
\end{equation}
 It also contains the 6-boson interaction term $g^2 
\phi^{a_{1}b_{1}c_{1}}\phi^{a_{1}b_{2}c_{2}}\phi^{a_{2}b_{1}c_{3}}\phi^{a_{2}b_{3}c_{1}} \phi^{a_{3}b_{2}c_{3}}\phi^{a_{3}b_{3}c_{2}} 
$, whose index structure is the same as that found in operator $O_{6}$ shown in (\ref{prism}); it can be represented graphically as
the prism (see figure \ref{Alloprs}). 

Such a construction may be viewed as a dimensional reduction of the ${\cal N}=1$ supersymmetric theory in $d=3$, where it is renormalizable (the field content is
a real scalar $\phi^{abc}$ and a two-component Majorana fermion $\chi^{abc}$). The interaction becomes relevant in $d<3$ so that the theory may flow to an interacting CFT.
Alternatively, we could study a renormalizable ${\cal N}=2$ supersymmetric theory in $d=3$, whose field content is a complex 
scalar $\phi^{abc}$ and a two-component Dirac fermion.
Since the $R$-charge of $\phi$ is fixed by the quartic superpotential (\ref{superpot}) to be $1/2$, we know that its exact dimension is 
\begin{align}
\Delta_\phi=\frac{d-1}{2} R=  \frac{d-1}{4}\ .
\end{align}
The dimension of the fermion superpartner is then $\Delta_\psi= \Delta_\phi+\frac 1 2= \frac{d+1}{4}$, so that the interaction term 
$\bar \psi^{a_{1}b_{1}c_{1}}\phi^{a_{1}b_{2}c_{2}}\psi^{a_{2}b_{1}c_{2}}\phi^{a_{2}b_{2}c_{1}} + {\rm c.\ c.}$ has dimension $d$.
The scalar potential
\begin{align}
V=  |g|^2
\phi^{a_{1}b_{1}c_{1}}\phi^{a_{1}b_{2}c_{2}}\phi^{a_{2}b_{1}c_{2}}\bar \phi^{a_{3}b_{3}c_{1}} \bar \phi^{a_{3}b_{2}c_{4}} \bar \phi^{a_{2}b_{3}c_{4}}
\end{align}
is, of course, non-negative. 
It would also be interesting to study a ``colored" supersymmetric theory with superfields $\Phi_A^{abc}$ and quartic  superpotential
\begin{align}
W=g
\Phi_0^{a b c }\Phi_1^{a d e}\Phi_2^{f b e} \Phi_3^{fdc}
\ .
\end{align}
The existence of perturbative expansion using supergraphs suggests that
the large $N$ limit is dominated by melonic diagrams. The quantum properties of these supersymmetric theories in $d<3$ may be studied using both the large $N$ Schwinger-Dyson equations
and the $3-\epsilon$ expansion. We hope to address these problems in the future.  

\bigskip
\bigskip
\centerline{\bf Note Added in Proof}
\bigskip

The construction of theories for a single rank $3$ tensor field with the quartic interaction 
(\ref{uncoloredint}) may be
generalized to a single rank $q-1$ tensor with the $O(N)^{q-1}$ symmetric interaction of order $q$. Since the indices of each $O(N)$ group must be contracted pairwise,
$q$ has to be even. Every pair of tensors in the interaction term has one index in common. For example, for $q=6$ the explicit form of the 
interaction of a real rank $5$ tensor is
\begin{align}
V_{\rm int}= i \frac {g}{6}
\psi^{a_1 b_1 c_1 d_1  e_1}\psi^{a_1 b_2 c_2 d_2 e_2} \psi^{a_2 b_2 c_3 d_3 e_1} 
\psi^{a_2 b_3 c_2 d_1 e_3}\psi^{a_3 b_3 c_1 d_3 e_2} \psi^{a_3 b_1 c_3 d_2 e_3}
\ .\notag
\end{align}
Such rank $q-1$ tensor theories have a large $N$ limit with $g^2 N^{(q-1)(q-2)/2}$ held fixed, which is dominated by the melonic diagrams.
When $\psi$ is taken to be a real anti-commuting rank $q-1$ tensor in $0+1$ dimensions, we find an $O(N)^{q-1}$ symmetric quantum mechanical model. It
provides a tensor implementation of the version of SYK model where the random interaction couples $q$ fermions.

\section*{Acknowledgments}

We are very grateful to E. Witten for important advice on many aspects of this paper and for really useful comments on a draft.
We thank R. Gurau for useful comments on a draft and especially for pointing out reference \cite{Carrozza:2015adg}.
We also thank S. Giombi, D. Gross, J. Maldacena, J. Murugan, A. Polyakov and D. Stanford for very useful discussions.
The work of IRK and GT was supported in part by the US NSF under Grant No.~PHY-1620059. GT acknowledges the support of a Myhrvold-Havranek Innovative Thinking Fellowship.


\bibliographystyle{ssg}
\bibliography{TenMod}

\end{document}